\documentclass[conference,a4paper]{APSIPA2018}
\usepackage{multirow}
\usepackage{graphicx}
\usepackage{amsmath}
\usepackage[psamsfonts]{amssymb}
\usepackage{amsxtra}
\usepackage{threeparttable}
\usepackage{array}
\usepackage{caption}
\usepackage{subfigure}

\begin{document}

\title{Double JPEG Compression Detection by Exploring the Correlations in DCT Domain}

\author{%
\authorblockN{%
Pengpeng Yang,\authorrefmark{1}\authorrefmark{2}
Rongrong Ni,\authorrefmark{1}\authorrefmark{2} and
Yao Zhao\authorrefmark{1}\authorrefmark{2}
}
\authorblockA{%
\authorrefmark{1}
Beijing Jiaotong University, Institute of Information Science, Beijing, China 
}
\authorblockA{%
\authorrefmark{2}
Beijing Key Laboratory of Advanced Information Science and Network Technology \\
E-mail: \{ppyang, rrni, yzhao\}@bjtu.edu.cn}
}

\maketitle
\thispagestyle{empty}

\begin{abstract}
In the field of digital image processing, JPEG image compression technique has been widely applied. And numerous image processing software suppose this. It is likely for the images undergoing double JPEG compression to be tampered. Therefore, double JPEG compression detection schemes can provide an important  clue for image forgery detection. In this paper,  we propose an effective  algorithm to detect double JPEG compression with  different  quality  factors.    Firstly, the quantized DCT coefficients  with  same frequency are  extracted to build the new data matrices. Then,  considering the direction effect  on the correlation  between the adjacent positions in  DCT domain, twelve kinds of high-pass filter templates with different  directions  are  executed  and the translation probability matrix is calculated for  each  filtered  data.  Furthermore,  principal component analysis and support vector machine technique are applied to reduce the feature dimension and train a classifier, respectively.  Experimental results have demonstrated that the proposed method is effective and has comparable performance. 
\end{abstract}

\section{Introduction}
With the development of digital image processing technology, enormous image processing software that free and easy to operate are spreading across the world. The authenticity and integrity of digital images have encountered huge challenges and seeing is no longer believing [1-3], which might be harmful to social public security. To overcome such challenges, digital image forensics  is put forward and has become a vital issue in recent years. 

As an important branch of digital image forensics, double JPEG image forensic [4,5] has attracted much attention in forensic community, which can provide a clue for studying the authenticity of digital images. According to the status of the quality factor during double compression test , the studies can be divided into two categories: $Q_{1}=Q_{2}$ ($Q_{1}$ and $Q_{2}$ denote the quality factor during the first and second compression, respectively), and $Q_{1}\neq Q_{2}$. Since the situation with $Q_{1}\neq Q_{2}$ is a common practice, some researchers have paid much attention to this problem and a number of algorithms have been proposed. An effective scheme relied on the shape of the histogram of the DCT coefficients [6,7]. The authors observed that double JPEG compression would induce periodic artifacts in the histogram of the DCT coefficients. In addition, Benford’s Law or the First-Digit phenomenon [8-11] was used to solve this issue. The authors proposed that the distribution of the first number of the DCT coefficients would be destroyed after double JPEG compression. Besides, Chen et. al [12] put forward a method based on machine learning. The difference matrices of the DCT coefficients along with four directions were obtained firstly. Then the translation probability matrices were calculated to describe the feature of the correlations in DCT domain. Recently, the data-driven methods [13-17] have been presented and achieved good performance. 

In this paper, focusing on the case of $Q_{1}\neq Q_{2}$, we propose an improving algorithm by exploring the correlations in DCT domain. Firstly, inspired by Chen's work, the different matrices of the DCT coefficients along with twelve kinds of the directions are calculated. In this way, the changes in correlations caused by double JPEG compression would be better described. Furthermore, considering the limitless of the global features, we extract twenty sub-matrices from the DCT coefficients according to the MODE (one MODE is at the same frequency). The proposed method is evaluated on a public image database: UCID [18] and compared with two previous algorithms. Experimental results prove its effectiveness.
The rest of this paper is organized as follow. Section 2 give a brief description of the JPEG compression. Section 3 present in detail the proposed method. experimental results are discussed in Section 4 and conclusion is drawn in Section 5.

\section{JPEG Compression}

\begin{figure}[t]
	\centering
	\includegraphics[width=8cm,height=5cm]{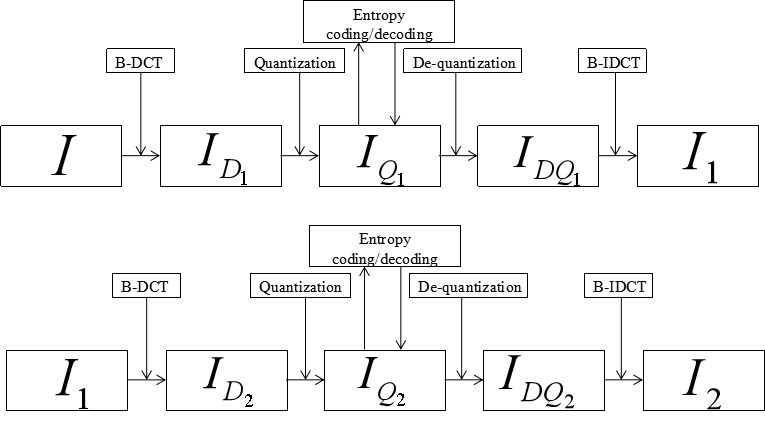}
	\caption{The workflow of single and double JPEG compression and decompression.}
\end{figure}

\begin{figure*}
	\centering
	\includegraphics[width=14cm,height=6cm]{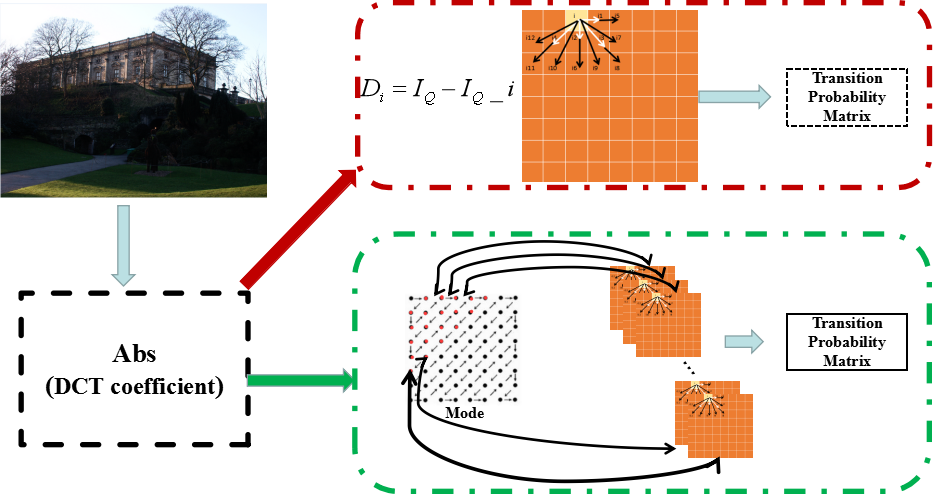}
	\caption{The overall framework of proposed feature extraction method. }
\end{figure*}

JPEG compression, a popular lossy compression scheme widely used in image data compression, was proposed by the Joint Photographic Experts Committee. There are six important steps in image compression and decompression system: B-DCT (Block Discrete Cosine Transform base on non-overlapping blocks of size 8x8 pixels), quantization, entropy coding, entropy decoding, de-quantization, and B-IDCT (Block Inverse Discrete Cosine Transform), which are executed in order.

The diagram of single and double JPEG compression and decompression is shown in Fig. 1. The procedure from $ I$ (the raw image) to $I_{Q_{1}}$ (the single compressed image) illustrates the single JPEG compression with $Q_{1}$ . The process from $I_{1}$ (decompressed data) to $I_{Q_{2}}$ denotes the double JPEG compression with $Q_{2}$. In the following sections, only one case, $Q_{1}\neq Q_{2}$, will be discussed, which seems to be a common practice.

\section{Proposed method}
The well-known machine learning technique represents an efficient tool for the detection of double JPEG compression. Inspired by Chen's machine learning-based scheme [12], the two-phase framework for feature extraction and classification is applied into our method. The flow-process for feature descriptors extraction is presented in Fig. 2, which includes two different ways as shown in red and green dotted boxes. For the red one, the correlations along various directions in DCT domain are introduced. Specifically, the twelve types of high-pass filters are executed on the quantized DCT coefficients and the translation probability matrix is calculated for each filtered data to establish the feature set. For the green one, in order to further improve performance, the correlations of the zig-zag sequence of firstly twenty AC coefficients are separately considered based on the above idea. In addition, principle component analysis (PCA) is used to reduce feature dimension. In the stage of feature classification, the SVM [19] classifiers with two degree polynomial kernel are trained through the above features. The description of feature extraction in detail is as follows.    

\begin{figure*}[t]
	\subfigure[]{                    
		\begin{minipage}{5.5cm}
			\centering                                                          
			\includegraphics[width=5cm,height=4.5cm]{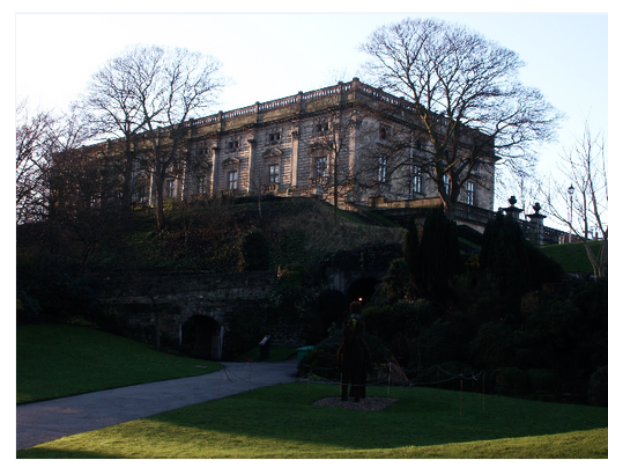}               
		\end{minipage}}
		\subfigure[]{                       
			\begin{minipage}{5.5cm}
				\centering                                                         \includegraphics[width=5cm,height=4.5cm]{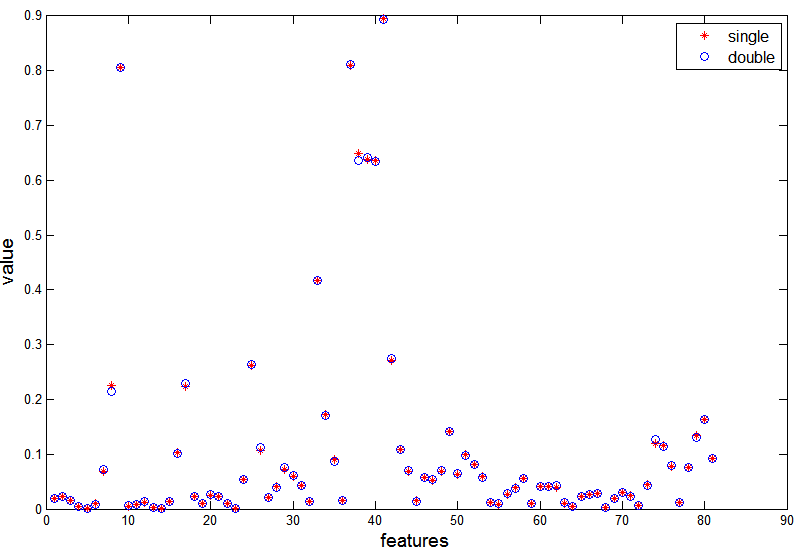}                
			\end{minipage}}
			\subfigure[]{                    
				\begin{minipage}{5.5cm}
					\centering                                                          
					\includegraphics[width=5cm,height=4.5cm]{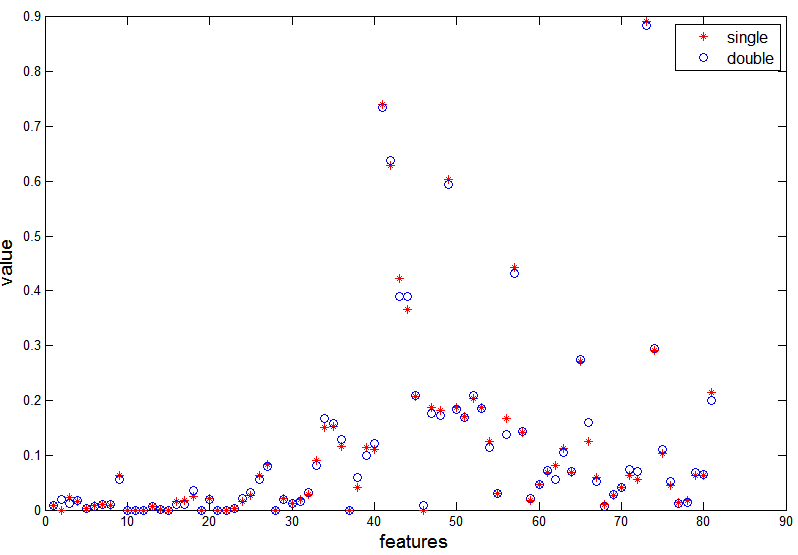}               
				\end{minipage}}
				
				\subfigure[]{                    
					\begin{minipage}{5.5cm}
						\centering                                                          
						\includegraphics[width=5cm,height=4.5cm]{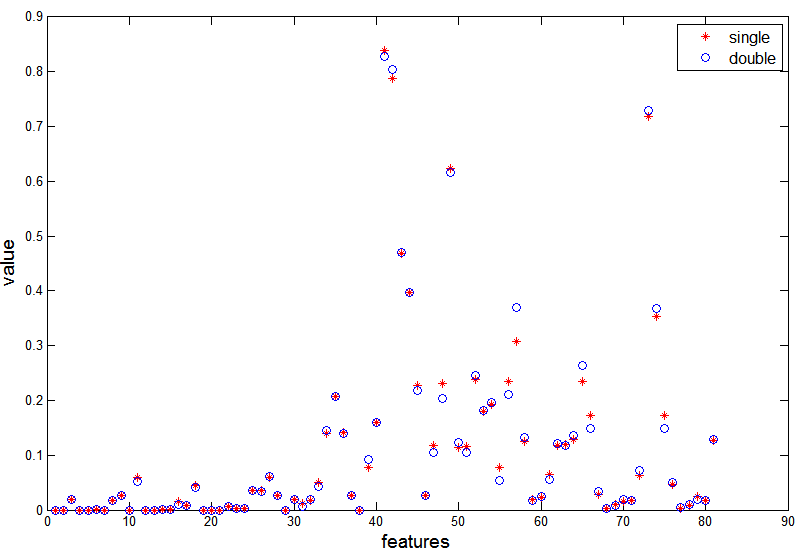}               
					\end{minipage}}
					\subfigure[]{                       
						\begin{minipage}{5.5cm}
							\centering                                                         \includegraphics[width=5cm,height=4.5cm]{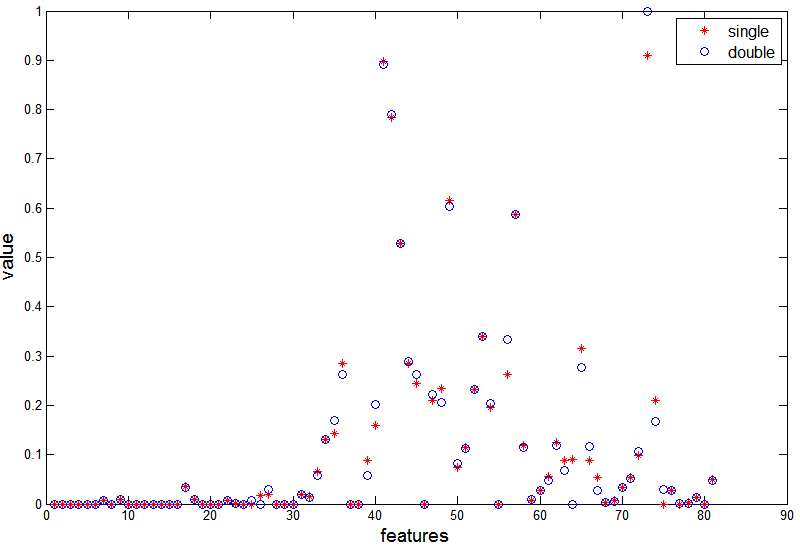}                
						\end{minipage}}
						\subfigure[]{                    
							\begin{minipage}{5.5cm}
								\centering                                                          
								\includegraphics[width=5cm,height=4.5cm]{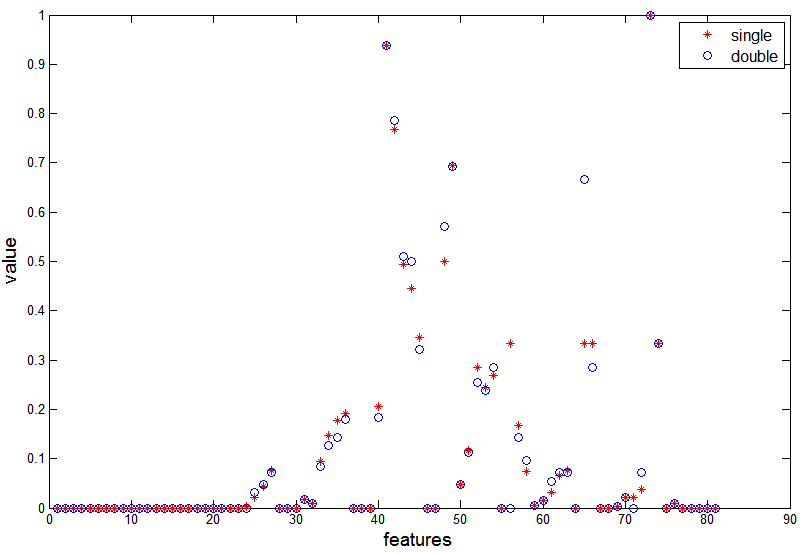}               
							\end{minipage}}		
							\caption{Visualization on the global and local features of single and double JPEG compressed image. The abscissa and ordinate denote the index and values of the feature, respectively. (a) shows the original image; (b) represents the feature of the correlations among multiple directions with $i=1$; the firstly four MODE features with $i=1$ are shown in (c),(d),(e),(f).}                         
							\label{fig:1}              
						\end{figure*}

\subsection{The Correlations Among Multiple Directions}
According to the previous work [12], there is no double that the correlations in DCT domain provide an important clue for double JPEG compression detection and the transition probability matrix can effectively describe the changes of the correlations. Following this way, the correlations among multiple directions are explored in this part as shown in the red dotted box of Fig. 2. Firstly, the quantized DCT coefficients for the luminance channel, $I_{Q}$, are obtained by jpeg\_read function in Matlab and taking its absolute values to build the quantized matrix. Then the twelve types of high-pass filters along with different directions are executed on the quantized matrix. The filtered data are truncated with a threshold.
\begin{equation}\label{key}
D_{i} = trun_{T}(|I_{Q}| - |I_{Q}\_i|) =  trun_{T}(-F_{i}*|I_{Q}|), i\epsilon \left \{ 1,2,3... 10,11,12 \right\}
\end{equation}  
where $|.|$ is the absolute value function, $I_{Q}\_i$ denotes the surrounding values of $I_{Q}$ as shown in Fig. 2 and $F_{i}$ represents the high-pass filter with 5x5 windows.

\begin{equation}\label{key}
F_{i} = \begin{bmatrix}
0 &  0&  0&  0& 0\\ 
0 &  0&  0&  0& 0\\ 
0 &  0&  -1&  1_{1}(i)& 1_{5}(i)\\ 
1_{12}(i) &  1_{4}(i)&  1_{2}(i)&  1_{3}(i)& 1_{7}(i)\\ 
1_{11}(i) &  1_{10}(i)&  1_{6}(i)&  1_{9}(i)& 1_{8}(i)
\end{bmatrix}
\end{equation}
where $1_{*}(i)$ is the indicator function, which means that $1_{*}(i)=1$ if and only if $i=*$, otherwise  $1_{*}(i)=0$. $trun_{T}()$ is truncate operation. In our experiments, the value for $T$ is setted to 4.
\begin{equation}\label{key}
trun_{T}(x) = \left\{\begin{matrix}
-T &, x< -T\\ 
x &  ,-T\leqslant  x\leqslant T \\ 
T &,x>T 
\end{matrix}\right.
\end{equation}
Next, the transition probability matrices for each filtered data are calculated as follows,
\begin{equation}\label{key}
P_{i}(p,q)=\frac{\sum \sum 1_{p,q}(D_{i}(m,n),D_{i}(\bar{m},\bar{n}))}{\sum \sum 1_{p}(D_{i}(m,n))},   \left \{p,q\right \}\epsilon \left [  -T,T\right ]  
\end{equation}

\begin{equation}\label{key}
1_{p,q}(D_{i}(m,n),D_{i}(\bar{m},\bar{n})) = \left\{\begin{matrix}
1 & ,D_{i}(m,n)=p\cap D_{i}(\bar{m},\bar{n})=q\\ 
0 &, others 
\end{matrix}\right.
\end{equation}
where $m,n,\bar{m},\bar{n}$, $p$, $q$ denote the index of matrix. $(m,n)$ and $(\bar{m},\bar{n})$ have the same position relationship with $I_{Q}$ and $I_{Q}\_{i}$. In the end, considering the symmetry of DCT basis functions and the sign symmetry of the translation matrices, $P_{i}(p,q)$ are combined as follows and the feature of correlations among multiple directions with 410 dimensions, $\bar{P_{i}}(\bar{p},\bar{q})$, would be generated.

\begin{equation}\label{key}
\begin{matrix}
\bar{P_{1}}(p,q) = (P_{1}(p,q)+P_{2}(p,q))/2\\ 
\bar{P_{5}}(p,q) = (P_{5}(p,q)+P_{6}(p,q))/2\\ 
\bar{P_{i}}(\bar{p},\bar{q}) = (P_{i}(\bar{p},\bar{q})+P_{i}(-\bar{p},-\bar{q}))/2\\
\left \{ \bar{p},\bar{q} \right \}\epsilon\left [ 0,T \right ]
\end{matrix}
\end{equation}

\subsection{The Correlations Based on MODE}
To some extent the above feature can represent the global information of the images. However, the local feature would provide a new perspective sometime. Therefore, the MODE is introduced to our method, which is supposed to improve performance. Note that one MODE is at the same frequency. The framework of feature extraction about the correlations based on MODE is shown in the green dotted box of Fig. 2. Differing from the above feature, same MODE in $I_{Q}$ is firstly extracted to construct the sub-matrices. It should be noted that only the firstly twenty MODEs in zig-zag order are used in this work. In order to verify the effectiveness of the feature based on MODE, we randomly chose one image from UCID dataset and extract its features of the quantized DCT coefficients and top four MODEs coefficients with $i=1$. The difference of the features between single and double JPEG compression image is visualized in Fig. 3. It can be seen that the MODE-based features make it easier to distinguish. Furthmore, the MODE-based feature has up to 17010 dimensions and PCA is applied to reduce it to 1300.

\section{Experiments}
In this Section, the experiments are conducted based on the public image database collected for image forensics, UCID [18] where 1338 raw images are included, to evaluate the effectiveness of the proposed method. These images are compressed with quality factors $Q_{1}$ and $Q_{2}$ in sequence to generate the single and double JPEG compressed images, where $\left \{ Q_{1},Q_{2} \right \}\epsilon \left \{ 50,55,60,65,70,75,80,85,90,95 \right \}$ and $Q_{1}\neq Q_{2}$. We randomly select 1138 single and double images as training data to train the SVM classifiers with two degree polynomial kernel. The others are assigned as testing data. The detection accuracy is averaged over 20 random experiments and the results are shown in the Tables below. 

One can see from Table I and III that the proposed algorithm based on the correlations among multiple directions has better performance comparing with the machine learning based method [12], which indicates that the idea of multiple directions introduced shall certainly be beneficial to the detection of double JPEG compression. However, detection accuracy of the above schemes is worse in the cases of $Q_{1}> Q_{2}$. The reason could be the limitation of the global feature, as we discussed in Section III.B. As shown in Table IV, using local feature of the correlations based on MODE typically receives a performance boost and obtains comparable detection performance with Benford's Law based method [9]. In addition, it can be seen that the proposed scheme based on MODE gets better performance in the cases of $Q_{1}=95$.

\section{Conclusions}
  In this paper, focusing on the double JPEG compression detection, we explore the features in DCT domain from two perspectives: the correlations between various directions and the correlations based on MODE. The proposed method by fusing the above points together obtains comparable performance with Benford's Law based scheme and has higher detection accuracy in much more difficult circumstances, $Q_{1}=95$.

\section*{Acknowledgment}

This work was supported in part by National NSF of China (61672090, 61332012), the National Key Research and Development of China (2016YFB0800404), Fundamental Research Funds for the Central Universities (2015JBZ002, 2017YJS054).

\begin{table*}[t!]
 	\centering
 	\renewcommand\arraystretch{1.3}
 	\caption{The detection accuracy of Machine Learning-based method [12].}
 	\label{my-label}
 	\begin{tabular}{|p{2cm}<{\centering} |p{1.5cm}<{\centering}|p{1cm}<{\centering}|p{1cm}<{\centering}|p{1cm}<{\centering}|p{1cm}<{\centering}|p{1cm}<{\centering}|p{1cm}<{\centering}|p{1cm}<{\centering}|p{1cm}<{\centering}|p{1cm}<{\centering}|}
 		\hline
 		\textbf{Q1\textbackslash{}Q2} & \textbf{50}  & \textbf{55}  & \textbf{60}  & \textbf{65}  & \textbf{70}  & \textbf{75}  & \textbf{80}  & \textbf{85}  & \textbf{90}  & \textbf{95}  \\ \hline
 		\textbf{50}                   & \textbf{－---}   & \textbf{97}  & \textbf{100} & \textbf{100} & \textbf{100} & \textbf{100} & \textbf{100} & \textbf{100} & \textbf{100} & \textbf{100} \\ \hline
 		\textbf{55}                   & \textbf{96}  & \textbf{－---}   & \textbf{99}  & \textbf{100} & \textbf{100} & \textbf{100} & \textbf{100} & \textbf{100} & \textbf{100} & \textbf{100} \\ \hline
 		\textbf{60}                   & \textbf{100} & \textbf{99}  & \textbf{－---}   & \textbf{99}  & \textbf{100} & \textbf{100} & \textbf{100} & \textbf{100} & \textbf{100} & \textbf{100} \\ \hline
 		\textbf{65}                   & \textbf{100} & \textbf{100} & \textbf{100} & \textbf{－---}   & \textbf{99}  & \textbf{100} & \textbf{100} & \textbf{100} & \textbf{100} & \textbf{100} \\ \hline
 		\textbf{70}                   & \textbf{100} & \textbf{100} & \textbf{100} & \textbf{100} & \textbf{－---}   & \textbf{100} & \textbf{100} & \textbf{100} & \textbf{100} & \textbf{100} \\ \hline
 		\textbf{75}                   & \textbf{99}  & \textbf{100} & \textbf{100} & \textbf{100} & \textbf{100}    & \textbf{－---}   & \textbf{100} & \textbf{100} & \textbf{100} & \textbf{100} \\ \hline
 		\textbf{80}                   & \textbf{100} & \textbf{100} & \textbf{99}  & \textbf{100} & \textbf{100} & \textbf{100} & \textbf{－---}   & \textbf{100} & \textbf{100} & \textbf{100} \\ \hline
 		\textbf{85}                   & \textbf{98}  & \textbf{89}  & \textbf{99}  & \textbf{100} & \textbf{99}  & \textbf{100} & \textbf{100} & \textbf{－---}   & \textbf{100} & \textbf{100} \\ \hline
 		\textbf{90}                   & \textbf{83}  & \textbf{97}  & \textbf{96}  & \textbf{99}  & \textbf{96}  & \textbf{100} & \textbf{99}  & \textbf{100} & \textbf{－---}   & \textbf{100} \\ \hline
 		\textbf{95}                   & \textbf{63}  & \textbf{62}  & \textbf{76}  & \textbf{74}  & \textbf{86}  & \textbf{77}  & \textbf{94}  & \textbf{85}  & \textbf{100} & \textbf{－---}   \\ \hline
 	\end{tabular}
 \end{table*}

 \begin{table*}[]
 	\centering
 	\renewcommand\arraystretch{1.3}
 	\caption{The detection accuracy of Benford's Law-based method [9].}
 	\label{my-label}
 	\begin{tabular}{|p{2cm}<{\centering} |p{1.5cm}<{\centering}|p{1cm}<{\centering}|p{1cm}<{\centering}|p{1cm}<{\centering}|p{1cm}<{\centering}|p{1cm}<{\centering}|p{1cm}<{\centering}|p{1cm}<{\centering}|p{1cm}<{\centering}|p{1cm}<{\centering}|}
 		\hline
 		\textbf{Q1/Q2} & \textbf{50}  & \textbf{55}  & \textbf{60}  & \textbf{65}  & \textbf{70}  & \textbf{75}  & \textbf{80}  & \textbf{85}  & \textbf{90}  & \textbf{95}  \\ \hline
 		\textbf{50}    & \textbf{－---}    & \textbf{98}  & \textbf{100} & \textbf{100} & \textbf{100} & \textbf{100} & \textbf{100} & \textbf{100} & \textbf{100} & \textbf{100} \\ \hline
 		\textbf{55}    & \textbf{98}  & \textbf{－---}   & \textbf{100} & \textbf{100} & \textbf{100} & \textbf{100} & \textbf{100} & \textbf{100} & \textbf{100} & \textbf{100} \\ \hline
 		\textbf{60}    & \textbf{100} & \textbf{99}  & \textbf{－---}   & \textbf{100} & \textbf{100} & \textbf{100} & \textbf{100} & \textbf{100} & \textbf{100} & \textbf{100} \\ \hline
 		\textbf{65}    & \textbf{100} & \textbf{100} & \textbf{99}  & \textbf{－---}   & \textbf{99}  & \textbf{100} & \textbf{100} & \textbf{100} & \textbf{100} & \textbf{100} \\ \hline
 		\textbf{70}    & \textbf{99}  & \textbf{100} & \textbf{100} & \textbf{100} & \textbf{－---}    & \textbf{99}  & \textbf{100} & \textbf{100} & \textbf{100} & \textbf{100} \\ \hline
 		\textbf{75}    & \textbf{95}  & \textbf{99}  & \textbf{100} & \textbf{100} & \textbf{100} & \textbf{－---}    & \textbf{100} & \textbf{100} & \textbf{100} & \textbf{100} \\ \hline
 		\textbf{80}    & \textbf{100} & \textbf{99}  & \textbf{100} & \textbf{100} & \textbf{100} & \textbf{100} & \textbf{－---}   & \textbf{100} & \textbf{100} & \textbf{100} \\ \hline
 		\textbf{85}    & \textbf{99}  & \textbf{98}  & \textbf{99}  & \textbf{99}  & \textbf{99}  & \textbf{100} & \textbf{100} & \textbf{－---}   & \textbf{100} & \textbf{100} \\ \hline
 		\textbf{90}    & \textbf{97}  & \textbf{98}  & \textbf{98}  & \textbf{98}  & \textbf{99}  & \textbf{99}  & \textbf{99}  & \textbf{100} & \textbf{－---}     & \textbf{100} \\ \hline
 		\textbf{95}    & \textbf{76}  & \textbf{80}  & \textbf{90}  & \textbf{95}  & \textbf{92}  & \textbf{96}  & \textbf{98}  & \textbf{99}  & \textbf{99}  & \textbf{－---}   \\ \hline
 	\end{tabular}
 \end{table*}

 \begin{table*}[]
 	\centering
 	\renewcommand\arraystretch{1.3}
 	\caption{The detection accuracy of proposed method based on the correlations among multiple directions}
 	\label{my-label}
 	\begin{tabular}{|p{2cm}<{\centering} |p{1.5cm}<{\centering}|p{1cm}<{\centering}|p{1cm}<{\centering}|p{1cm}<{\centering}|p{1cm}<{\centering}|p{1cm}<{\centering}|p{1cm}<{\centering}|p{1cm}<{\centering}|p{1cm}<{\centering}|p{1cm}<{\centering}|}
 		\hline
 		\textbf{Q1/Q2} & \textbf{50}  & \textbf{55}  & \textbf{60}  & \textbf{65}  & \textbf{70}  & \textbf{75}  & \textbf{80}  & \textbf{85}  & \textbf{90}  & \textbf{95}  \\ \hline
 		\textbf{50}    & \textbf{－---}   & \textbf{97}  & \textbf{100} & \textbf{100} & \textbf{100} & \textbf{100} & \textbf{100} & \textbf{100} & \textbf{100} & \textbf{100} \\ \hline
 		\textbf{55}    & \textbf{96}  & \textbf{－---}   & \textbf{99}  & \textbf{100} & \textbf{100} & \textbf{100} & \textbf{100} & \textbf{100} & \textbf{100} & \textbf{100} \\ \hline
 		\textbf{60}    & \textbf{100} & \textbf{99}  & \textbf{－---}   & \textbf{99}  & \textbf{100} & \textbf{100} & \textbf{100} & \textbf{100} & \textbf{100} & \textbf{100} \\ \hline
 		\textbf{65}    & \textbf{100} & \textbf{100} & \textbf{100} & \textbf{－---}   & \textbf{99}  & \textbf{100} & \textbf{100} & \textbf{100} & \textbf{100} & \textbf{100} \\ \hline
 		\textbf{70}    & \textbf{100} & \textbf{100} & \textbf{100} & \textbf{100} & \textbf{－---}   & \textbf{100} & \textbf{100} & \textbf{100} & \textbf{100} & \textbf{100} \\ \hline
 		\textbf{75}    & \textbf{99}  & \textbf{100} & \textbf{100} & \textbf{100} & \textbf{100} & \textbf{－---}   & \textbf{100}    & \textbf{100} & \textbf{100} & \textbf{100} \\ \hline
 		\textbf{80}    & \textbf{100} & \textbf{100} & \textbf{99}  & \textbf{100} & \textbf{100} & \textbf{100} & \textbf{－---}   & \textbf{100} & \textbf{100} & \textbf{100} \\ \hline
 		\textbf{85}    & \textbf{99}  & \textbf{96}  & \textbf{100} & \textbf{100} & \textbf{99}  & \textbf{100} & \textbf{100} & \textbf{－---}   & \textbf{100} & \textbf{100} \\ \hline
 		\textbf{90}    & \textbf{91}  & \textbf{98}  & \textbf{97}  & \textbf{99}  & \textbf{97}  & \textbf{100} & \textbf{99}  & \textbf{100} & \textbf{－---}   & \textbf{100} \\ \hline
 		\textbf{95}    & \textbf{72}  & \textbf{79}  & \textbf{83}  & \textbf{86}  & \textbf{92}  & \textbf{88}  & \textbf{96}  & \textbf{94}  & \textbf{100} & \textbf{－---}   \\ \hline
 	\end{tabular}
 \end{table*}

 \begin{table*}[t!]
 	\centering
 	\renewcommand\arraystretch{1.3}
 	\caption{The detection accuracy of proposed method based on MODE}
 	\label{my-label}
 	\begin{tabular}{|p{2cm}<{\centering} |p{1.5cm}<{\centering}|p{1cm}<{\centering}|p{1cm}<{\centering}|p{1cm}<{\centering}|p{1cm}<{\centering}|p{1cm}<{\centering}|p{1cm}<{\centering}|p{1cm}<{\centering}|p{1cm}<{\centering}|p{1cm}<{\centering}|}
 		\hline
 		\textbf{Q1/Q2} & \textbf{50}  & \textbf{55}  & \textbf{60}  & \textbf{65}  & \textbf{70}  & \textbf{75}  & \textbf{80}  & \textbf{85}  & \textbf{90}  & \textbf{95}  \\ \hline
 		\textbf{50}    & \textbf{－---}    & \textbf{95}  & \textbf{100} & \textbf{100} & \textbf{100} & \textbf{100} & \textbf{100} & \textbf{100} & \textbf{100} & \textbf{100} \\ \hline
 		\textbf{55}    & \textbf{94}  & \textbf{－---}   & \textbf{99}  & \textbf{100} & \textbf{100} & \textbf{100} & \textbf{100} & \textbf{100} & \textbf{100} & \textbf{100} \\ \hline
 		\textbf{60}    & \textbf{100} & \textbf{99}  & \textbf{－---}   & \textbf{99}  & \textbf{100} & \textbf{100} & \textbf{100} & \textbf{100} & \textbf{100} & \textbf{100} \\ \hline
 		\textbf{65}    & \textbf{100} & \textbf{100} & \textbf{99}  & \textbf{－---}   & \textbf{100} & \textbf{100} & \textbf{100} & \textbf{100} & \textbf{100} & \textbf{100} \\ \hline
 		\textbf{70}    & \textbf{100} & \textbf{100} & \textbf{100} & \textbf{99}  & \textbf{－---}    & \textbf{100} & \textbf{100} & \textbf{100} & \textbf{100} & \textbf{100} \\ \hline
 		\textbf{75}    & \textbf{99}  & \textbf{100} & \textbf{100} & \textbf{100} & \textbf{100} & \textbf{－---}   & \textbf{100} & \textbf{100} & \textbf{100} & \textbf{100} \\ \hline
 		\textbf{80}    & \textbf{99}  & \textbf{99}  & \textbf{99}  & \textbf{100} & \textbf{100} & \textbf{100} & \textbf{－---}   & \textbf{100} & \textbf{100} & \textbf{100} \\ \hline
 		\textbf{85}    & \textbf{98}  & \textbf{97}  & \textbf{99}  & \textbf{99}  & \textbf{99}  & \textbf{100} & \textbf{100} & \textbf{－---}   & \textbf{100} & \textbf{100} \\ \hline
 		\textbf{90}    & \textbf{94}  & \textbf{97}  & \textbf{99}  & \textbf{98}  & \textbf{98}  & \textbf{99}  & \textbf{99}  & \textbf{100} & \textbf{－---}   & \textbf{100} \\ \hline
 		\textbf{95}    & \textbf{79}  & \textbf{86}  & \textbf{92}  & \textbf{94}  & \textbf{96}  & \textbf{98}  & \textbf{98}  & \textbf{99}  & \textbf{100} & \textbf{－---}   \\ \hline
 	\end{tabular}
 \end{table*}







\end{document}